\documentclass[aps,prb,floats,twocolumn]{revtex4}
\usepackage[dvips]{graphicx}
\usepackage{amsmath}

\begin{document}

\title{Low density approach to the Kondo-lattice model}
\author{
W. Nolting$^{a}$,\ G. G. Reddy$^{b}$,\ A. Ramakanth$^{b}$\ and D. Meyer$^{c}$}
\affiliation{$^{a}$Humboldt-Universit\"at zu Berlin, Institut f\"ur Physik,
 Lehrstuhl Festk\"{o}rpertheorie,Invalidenstr. 110, 110115 Berlin, Germany\\
 $^{b}$KakatiyaUniversity, Department of Physics, Warangal-506009, India\\
 $^{c}$Department of Mathematics, Imperial College, 180 Queen's Gate, London SW7 2BZ, United Kingdom}
\date{\today}

\begin{abstract}

We propose a new approach to the (ferromagnetic) Kondo-lattice model in the low density 
region, where the model is thought to give a reasonable frame work for manganites with perovskite 
structure exhibiting the "colossal magnetoresistance" -effect. Results for the temperature-
dependent quasiparticle density of states are presented. Typical features can be interpreted in 
terms of elementary spin-exchange processes between itinerant conduction electrons and 
localized moments. The approach is exact in the zero bandwidth limit for all temperatures and 
at $T=0$ for arbitrary bandwidths, fulfills exact high-energy expansions and reproduces 
correctly second order perturbation theory in the exchange coupling.
\end{abstract}


\maketitle

\section{Introduction}
The Kondo-lattice model (KLM) [1] describes the interplay of itinerant electrons in a partially 
filled energy band with quantum mechanical spins (magnetic moments), localized at certain 
lattice sites. Characteristic model properties result from an interband exchange interaction 
between the two subsystems.

On the one hand, the energy bandstructure is modified by the magnetic state of the spin system 
(temperature dependences, band splittings, band deformations), while, on the other, the 
magnetic state of the spin system is even provoked by the itinerant electrons because the KLM 
does not incorporate a direct exchange between the moments. The model-Hamiltonian consists of 
two parts
\begin{equation}
H=H_{s}+H_{sf}
\end{equation}
$H_{s}$ is the kinetic energy of itinerant band electrons,
\begin{equation}
H_{s}=\sum_{ij\sigma}T_{ij}c_{i\sigma}^{\dag}c_{j\sigma}
\end{equation} 
where $c_{i\sigma}^{\dag}(c_{i\sigma})$  is the creation (annihilation) operator of a band 
electron specified by the lower indices. $T_{ij}$ are the hopping integrals. The second term in (1) 
is an interband exchange term with coupling strength $J$, written as an intra-atomic interaction 
between the conduction electron 
spin $\mathbf{\sigma}_{i}$ and the localized magnetic moment represented by the spin operator
 $\mathbf{S}_{i}$:
\begin{equation}
H_{sf}=-J\sum_{i}\mathbf{\sigma}_{i}\cdot \mathbf{S}_{i}.
\end{equation}
According to the sign of the exchange coupling $J$, a parallel ($J > 0$) or an antiparallel
 ($J < 0$) alignment of itinerant and localized spin is favoured with remarkable differences in 
 the physical properties. The parallel ($J > 0$) orientation is often referred to as 
 "ferromagnetic Kondo-lattice model" (FKLM); alternatively known as "s-f or s-d model". The 
 applications of the KLM are rather manifold.
 
\subsection{ Magnetic semiconductors:}

Prototypes are the Europium chalcogenides EuX (X = O, S, Se, Te) [2] which are known to 
exhibit a spectacular temperature dependence of the band states. The "red shift" of the optical 
absorption edge upon cooling from $T=T_{c}$ to $T=0K$ [2, 3] is due to a corresponding shift of the 
lower conduction band edge. There is clear evidence that in these materials the exchange J is 
positive, typically of order some tenth of eV. The coupling can therefore be classified as weak 
to intermediate. 

\subsection{ Semimagnetic semiconductors:}

In systems like $Cd_{1-x}Mn_{x}Te$ and $Hg_{1-x}Fe_{x}Se$ randomly distributed $Mn^{2+}$ or
 $Fe^{2+}$ ions provide localized magnetic moments which influence, via the exchange mechanism J,
 the band states of the II-VI semiconductors CdTe and HgSe. For moderate doping $x$, the moments 
 do not order collectively so that a striking temperature dependence, as that of the magnetic 
 semiconductors (EuX), cannot be expected. However, an anomalous magnetic field dependence of 
 optical transitions and therewith of the bandstructure is observed [4] ("giant Zeeman 
 splitting"). From respective experimental data, $J > 0$ can be concluded. The coupling must be
 classified as weak.

\subsection{Local-moment metals:}

In ferromagnetic metals such as the Rare Earth element Gd, the magnetism is due to strictly 
localized 4f electrons while the conductivity properties are determined by itinerant (5d, 6s) 
electrons. The $T=0$-moment of Gd is found to be $7.63\mu_{B}$ [5]. $7\mu_{B}$ stem from the 
exactly half-filled 4f shell. The excess-moment of $0.63\mu_{B}$ originates from an induced spin
 polarization 
of the "a priori" non-magnetic conduction bands, indicating a weak or intermediate coupling 
$J > 0$ [6]. Many of the recent research activities have been focussed on the temperature 
dependence of the induced exchange splitting. Is it collapsing for $T\rightarrow T_{c}$ or does
 it persist even in the paramagnetic phase [6, 7]? The $J$-induced correlation and quasi
particle effects in the valence and conduction bands of Gd (or equivalently Dy, Tb) lead to 
 highly complex and therefore controversially discussed photoemission data [8, 9], the 
 interpretation of which is far from settled (see review in [7]).While the magnetic ordering of
  the semiconductors and insulators 
(class 1) has to be explained via special superexchange mechanisms, which is beyond the field of 
application of the FKLM, it is commonly accepted that the collective magnetism of the "local
 moment" metals 
is caused by the RKKY interaction. The latter is also based on the exchange interaction $J$. 
The FKLM therefore provides, at least in a qualitative manner, a selfconsistent description
 of magnetic and electronic properties of materials such as Gd [6, 10].
 
\subsection{Manganites-perovskites}

Since the discovery of the "colossal magnetoresistance (CMR)" [11, 12], the manganese 
oxides with perovskite structures $T_{1-x}D_{x}MnO_{3}$ (T = La, Pr, Nd ; D = Sr, Ca, Ba, Pb)
 have attracted high scientific interest. The prototypes $La_{1-x}(Ca, Sr)_{x}MnO_{3}$ 
are since long the protagonists of the "double exchange" mechanism [13]. Replacing, in 
$La^{3+}Mn^{3+}O_{3}$, a trivalent $La^{3+}$ ion by a divalent earth-alkali ion ($Ca^{2+},
Sr^{2+}$) requires an additional electron from the manganese for the binding. The result is a 
homogenuous valence mixture of the manganese ion $(Mn_{1-x}^{3+}Mn_{x}^{4+})$ . The three 
$3d-t_{2g}$ electrons of $Mn^{4+}$ are considered as more or less localized 
forming a local $S = 3/2$ spin. The fourth electron in $Mn^{3+}$ is of $3d-e_{g}$ type and is
 itinerant. It is assumed that it interacts via intra-shell Hund's rule coupling ("double
  exchange model" [14]) with the $S =3/2$ spins. The manganites are bad electrical 
conductors. It has therefore to be assumed that the intraatomic coupling $J > 0$ is much 
stronger than the hopping-matrix element $|t|\ (J >> |t|)$. Theoretical estimates for the
 bandwidth yield $W = 1-2\ eV$ [15-17], experimental data propose $W = 3-4\ eV$ [18, 19]. 
 The exchange coupling $J$ is not very well known, the $J =1\ eV$ of refs. [15, 20] are 
 sometimes questioned as being too small [21]. In any case, the 
manganites belong to the strongly coupled FKLM which cannot be treated perturbatively 
with respect to $J$. The FKLM will certainly be overcharged to reproduce all the details 
of the rich phase diagram of $La_{1-x}Ca_{x}MnO_{3}$, e.g., according to which the ground state
 is antiferromagnetic for $x =0$ and 1, ferromagnetic for $x\approx 0.2-0.4$, with paramagnetic 
 regions and phase separations in between [12]. Nevertheless, the FKLM is thought to give a 
reasonable frame work for an at least qualitative understanding of the interesting physics of the 
manganites [22, 23]

\subsection{Heavy Fermions}

The above subclasses are all characterized by a ferromagnetic exchange interaction $J > 0$. The 
original Kondo-lattice model [24], however, refers to $J < 0$, favouring an antiparallel 
alignment of conduction electron spin and localized spin. This situation is obviously realized 
in the Heavy-Fermion systems, which are to be found especially among Ce-compounds and 
which have provoked intensive research activities because of their extraordinary physical 
properties. Doniach [24] was the first to point out that there should be a phase transition from 
a magnetic state for small $|J|$ to a non-magnetic Kondo state for large $|J|$ characterized by a 
screening of the local moments by the conduction electron spins. The magnetic state is due to the 
RKKY-interaction, which as an effect of second order $(\sim J^{2})$ is independent of the sign 
of $J$. However, the Kondo screening is of course absent for $J > 0$, i.e., for all 
the above discussed subclasses. For most of the Heavy-Fermion systems, the RKKY coupling favours an 
antiferromagnetic ordering of the local moments. In $CeCu_{6-x}Au_{x}$ the competitive behaviour 
of the RKKY and the Kondo screening tendencies can impressively be observed 
by varying the 
concentration $x$ [25]. $CeCu_{6}(x=0)$ is non-magnetic because of perfect Kondo screening, 
while for $x > 0.1$ the RKKY component dominates taking care for an antiferromagnetic 
ordering up to $x=1\ (CeCu_{5}Au)$ with increasing Neel-temperature $T_{N}$ for increasing $x$.

$J < 0$ does not necessarily lead to antiferromagnetism. The compound $CeSi_{x}$ is ferromagnetic 
for $1.6\le x\le 1.85$ [26] with a strongly reduced magnetic moment. The Curie temperature of 
the ferromagnetic Kondo system $(J < 0)$ $CeNi_{x}Pt_{1-x}$ first increases in between $x=0$ and 
$x = 0.5$ from 5.8K $(x=0)$ to about 8.6K $(x=0.5)$ in order to decrease then rapidly and 
disappearing eventually at $x=0.55$ [27]. The magnetic moment per Ce ion
 diminishes steadily with $x$ because of increasing Kondo screening and disappears completely 
 at $x = 0.95$.
 
  The above-presented list documents the rich variety of applications for 
 the KLM. Since the many-body problem of the Hamiltonian (1) could not be solved exactly up to now, 
approximations must be tolerated. Most of the recent theoretical papers, aiming at the CMR-materials, assume classical spins $S \rightarrow \infty$ [28-30], mainly in order to be able to
 apply "dynamical mean field theory" (DMFT) to the FKLM problem. The merits of DMFT, e. g., 
with respect to the Hubbard model, are indisputable, but the assumption of classical spins in 
the KLM appears very problematic. Several important features, as, e.g., the magnon emission and 
absorption by the itinerant electrons, are excluded from the very beginning. The importance of 
such effects has been discussed in detail in ref. [10]. Conclusions such
 as, that at $T=0$ the spins of the $e_{g}$ electrons are oriented parallel to the $t_{2g}-$spins [30],
  are correct only for $S \rightarrow \infty$. For any finite spin, there is a considerable amount 
  of $\downarrow-$spectral   weight overlapping with $\uparrow-$states even for very large $J$. 
  Recently, a DMFT-based approach to the KLM with quantum spins has been proposed [31] which uses 
  a fermionization of the local spin
  operators. The theory is restricted to $S=1/2$ but retains the quantum nature of the spins.
   A band splitting, which occurs already for relatively low interaction strengths, can be related
    to distinct elementary excitations, namely magnon emission and absorption by the itinerant
     electron and the formation of magnetic polarons. The results, which are in remarkable 
agreement with those from the "moment conserving decoupling approach" (MCDA) in ref.
 [10], confirm the importance of the quantum nature of the spins.

Due to some reasons, the above-mentioned theories [10, 31] are best justified for weak and 
intermediate couplings $J$. In this paper, we propose an approximate scheme which mainly 
aims at the strong coupling regime ($JS >> W$ : $W$ is the bandwidth) being nevertheless 
perturbationally correct up to order $J^{2}$. The idea is to construct a selfenergy ansatz which 
interpolates between exactly known limiting cases and reproduces the correct high-energy expansion
 of the selfenergy. To demonstrate the method as clearly as possible, we restrict our 
 considerations to the low-concentration region, performing the detailed calculation for a single 
 electron in an otherwise empty conduction band. The theory is outlined in Sect. II, while 
 Sect. III brings a discussion of the results.

\section{Theory}

\subsection{The many body problem}

The model-Hamiltonian (1) defines a non-trivial many-body problem, the exact solution of 
which is known only for a small number of special cases. For practical reasons, it is sometimes 
more convenient to use the second quantized form of the exchange interaction (3):
\begin{equation}
H_{sf}=-\frac{1}{2}J\sum_{j\sigma}\left(z_{\sigma}S_{j}^{z}n_{j\sigma}+S_{j}^{-\sigma}c_{j-\sigma}
^{\dag}c_{j\sigma}\right).
\end{equation}
Here we have used the abbreviations:
\begin{equation}
n_{j\sigma}=c_{j\sigma}^{\dag}c_{j\sigma}\ ;\ z_{\sigma}=\delta_{\sigma
\uparrow}-\delta_{\sigma\downarrow}\ ;\ S_{j}^{\sigma}=S_{j}^{x}+iz_
{\sigma}S_{j}^{y}.
\end{equation} 		
The first term in (4) describes an Ising-like interaction between the z-components of the 
localized and the itinerant spins. The second term refers to spin exchange processes between 
the two subsystems.

If we are mainly interested in the conduction electron properties, then the single-electron Green 
function,
\begin{equation}
G_{ij\sigma}(E)=\left<\left<c_{i\sigma};c_{j\sigma}^{\dag}\right>\right>_{E},
\end{equation}
is of primary interest. Its equation of motion reads:
\begin{equation}
\begin{split} 
\sum_{m}\left(E\delta_{im}-T_{im}\right)G_{mj\sigma}(E)=\hbar\delta_{ij}\hspace{3cm}\\
-\frac{1}{2}J\left(z_
{\sigma}I_{ii,j\sigma}(E)+F_{ii,j\sigma}(E)\right)
\end{split}
\end{equation}
where the two types of interaction terms in (4) lead to the "spinflip function ", 
\begin{equation}
F_{im,j\sigma}(E)=\left<\left<S_{i}^{-\sigma}c_{m-\sigma};c_{j\sigma}^{\dag}\right>\right>_{E}
\end{equation}
and the "Ising function":
\begin{equation}
I_{im,j\sigma}(E)=\left<\left<S_{i}^{z}c_{m\sigma};c_{j\sigma}^{\dag}\right>\right>_{E}.
\end{equation}
The two "higher" Green functions on the right-hand side of (7) prevent a direct solution of 
the equation of motion. A formal solution for the Fourier-transformed single-electron Green 
function,
\begin{equation}
G_{\mathbf{k}\sigma}(E)=\left<\left<c_{\mathbf{k}\sigma};c_{\mathbf{k}\sigma}^{\dag}\right>\right>_
{E}=\frac{\hbar}{E-\epsilon(\mathbf{k})-\Sigma_{\mathbf{k}\sigma}(E)}
\end{equation}
defines the in general complex self-energy $\Sigma_{\mathbf{k}\sigma}(E)$ by the ansatz
\begin{equation}
\left<\left<\left[H_{sf},c_{\mathbf{k}\sigma}\right]_{-};c_{\mathbf{k}\sigma}^{\dag}\right>\right>_
{E}=\Sigma_{\mathbf{k}\sigma}(E)G_{\mathbf{k}\sigma}(E).
\end{equation}
$\epsilon(\mathbf{k})$ are the Bloch energies:
\begin{equation}
\epsilon(\mathbf{k})=\frac{1}{N}\sum_{i,j}T_{ij}e^{i\mathbf{k}\cdot(\mathbf{R}_{i}-\mathbf{R}_{j})}.
\end{equation}
An illustrative  quantity which we are going to discuss in the following is the quasiparticle 
density of states (Q-DOS):
\begin{equation}
\rho_{\sigma}(E)=-\frac{1}{\hbar\pi N}\sum_{\mathbf{k}}Im G_{\mathbf{k}\sigma}\left(E+i0^{+}
\right).
\end{equation}
For the general case neither $\Sigma_{\mathbf{k}\sigma}(E)$ nor $G_{\mathbf{k}\sigma}(E)$ can be 
determined exactly. However, some rigorous statements are possible and shall now be listed up.

\subsection{Zero-bandwidth limit}

The final goal of our study is to arrive at a self-energy formula being credible first of all in
 the strong coupling limit $(JS >> W)$. That means, in particular, that our approach has to fulfill 
 the exactly solvable zero-bandwidth case [32]:
\begin{equation}
T_{ij}\rightarrow T_{0}\delta_{ij}\hspace{0.2cm};\hspace{0.2cm}\epsilon(\mathbf{k})\rightarrow 
T_{0}\hspace{0.2cm}\forall\mathbf{k}.
\end{equation}
The conduction band is shrunk to an N-fold degenerate level $T_{0}$. The localized spin system, 
however, is furtheron considered as collectively ordered for $T < T_{c}$ by any direct or indirect exchange 
interaction. The latter is not a part of the KLM. The localized magnetization $\left<S^{z}\right>$ 
Therefore enters the calculation as external parameter. With (14), the hierarchy of equations of 
motion for the single-electron Green function $G_{ij}(E)$, following from eq. (7), decouples 
exactly [32]. The result is a four-pole function:
\begin{equation}
G_{ii\sigma}^{(W=0)}(E)=\sum_{j=1}^{4}\frac{\alpha_{j\sigma}}{E-E_{j\sigma}}
\end{equation}
with spin-independent poles at
\begin{equation}
E_{1\sigma}=T_{0}-\frac{1}{2}JS\hspace{0.1cm};\hspace{0.1cm}E_{2\sigma}=T_{0}+\frac{1}{2}J(S+1).
\end{equation}
\begin{equation}
E_{3\sigma}=T_{0}-\frac{1}{2}J(S+1)\hspace{0.1cm};\hspace{0.1cm}E_{4\sigma}=T_{0}+\frac{1}{2}JS
\end{equation}
The $j=1,2$-excitations (Eq. (16)) refer to singly occupied sites; more strictly, they appear when the
 test electron is brought to a site, where no other "conduction" electron is present. It then 
 orients its spin parallel $(E_{1\sigma})$ or antiparallel $(E_{2\sigma})$ to the local
  spin. These excitations are bound to spin-dependent spectral weights:
\begin{equation}
  \begin{split}
\alpha_{1\sigma}=\frac{1}{2S+1}\left\{S+1+m_{\sigma}+\Delta_{-\sigma}\right .\\
\left .-(S+1)\left<n_{-\sigma}\right>\right\}
  \end{split}
\end{equation}
\begin{equation}
\alpha_{2\sigma}=\frac{1}{2S+1}\left\{S-m_{\sigma}-\Delta_{-\sigma}-S\left<n_{-\sigma}\right>.
\right\}
\end{equation}
Here we have abbreviated:
\begin{equation}
m_{\sigma}=z_{\sigma}\left<S^{z}\right>
\end{equation}
\begin{equation}
\Delta_{\sigma}=\left<S_{i}^{\sigma}c_{i-\sigma}^{\dag}c_{i\sigma}\right>+z_{\sigma}\left<S_{i}
^{z}n_{i\sigma}\right>
\end{equation}
The "mixed" correlation function $\Delta_{\sigma}$ can be derived via the spectral theorem from the 
Ising- and the spinflip-functions (8 and 9). Exploiting the equation of motion (7), this can even be 
expressed in terms of the single-electron Green function:
\begin{equation}
\Delta_{\sigma}=-\frac{1}{\pi\hbar}\frac{1}{N}\sum_{\mathbf{k}}\int_{-\infty}^{+\infty}dEf_{-}(E)
\left(E-\epsilon(\mathbf{k})\right)Im G_{\mathbf{k}\sigma}(E)
\end{equation}
where $f_{-}(E)=(1+e^{(E-\mu)})^{-1}$ is the Fermi function ($\mu$ is the chemical potential). 
Similarly it holds for the spin-dependent particle numbers: 
\begin{equation}
\left<n_{\sigma}\right>=-\frac{1}{\pi\hbar}\frac{1}{N}\sum_{\mathbf{k}}\int_{-\infty}^{+\infty}dE
f_{-}(E)Im G_{\mathbf{k}\sigma}(E).
\end{equation}
The expectation values in the spectral weights $\alpha_{1,2\sigma}$ are, therefore, all 
selfconsistently determinable by the required single-electron Green function itself.

 The two other poles $E_{3\sigma}$ and$E_{4\sigma}$ are bound to double occupancies of the lattice
  site. The test electron enters a site which is  already occupied by another electron with 
  opposite spin. The corresponding spectral weights,
\begin{equation}
\alpha_{3\sigma}=\frac{1}{2S+1}\left\{S\left<n_{-\sigma}\right>-\Delta_{-\sigma}\right\}
\end{equation}
\begin{equation}
\alpha_{4\sigma}=\frac{1}{2S+1}\left\{(S+1)\left<n_{-\sigma}\right>+\Delta_{-\sigma}\right\}
\end{equation}
vanish in the limit of zero band occupation. It may be considered a shortcoming of the KLM that
 the excitation energies (17) do not contain the Coulomb interaction energy. Switching on a
 Hubbard interaction U leads to an additive term U in $E_{3\sigma}$ as well as in $E_{4\sigma}$
 [32], shifting these excitations to higher energies. While the Hubbard-U is of course the exact 
 ansatz in the zero-bandwidth limit, it is not so obvious by which type of Coulomb interaction 
 the KLM should be extended ("correlated KLM" [30]) when aiming at one of the subclasses described
  in the Introduction. To avoid this ambiguity we restrict our following considerations to the
low-density limit $(n\rightarrow 0)$, where the selfenergy of the zero-bandwidth KLM reads 
according to eqs. (16)-(19): 
\begin{equation}
\Sigma_{\sigma}^{(W=0)}(E)\stackrel{n\rightarrow 0}{\longrightarrow}
 \frac{\frac{1}{4}J^{2}S(S+1)-\frac{1}{2}Jm_{\sigma}(E-T_{0})}
{E-T_{0}-\frac{1}{2}J(m_{\sigma}+1)}.
\end{equation}
This rigorous result will be exploited later for testing our approximate theory.

\subsection{Ferromagnetically saturated semiconductor}

There is another very instructive limiting case that can be treated exactly. It concerns a single 
electron in an otherwise empty conduction band interacting with a ferromagnetically saturated 
local moment system $(T=0)$. In the zero-bandwidth limit (Sect. II.2) for the $\uparrow$-spectrum, 
all the spectral weights (19), (24) and (25) disappear, except for $\alpha_{1\uparrow}=1$. In the 
$\downarrow$-spectrum the levels $E_{1\downarrow}$ and $E_{2\downarrow}$ survive with the weights
$\alpha_{1\downarrow}=\frac{1}{2S+1}$ and $\alpha_{2\downarrow}=\frac{2S}{2S+1}$.

For finite bandwidth, the mentioned special case is that of a ferromagnetically saturated 
semiconductor (EuO at $T=0$!) [10, 31, 33-35]. In this situation, an $\uparrow$-electron has no 
chance for a spin-flip, the corresponding quasiparticle density of states, $\rho_{\uparrow}(E)$, 
is therefore only rigidly shifted compared to the "free" DOS [10] and the self-energy is a
constant:
\begin{equation}
\Sigma_{\mathbf{k}\uparrow}^{(T=0,n=0)}(E)=\Sigma_{\uparrow}^{(T=0,n=0)}(E)=-\frac{1}{2}JS.
\end{equation}
The $\downarrow$-spectrum is more complicated since a $\downarrow$-electron has several 
possibilities to exchange its spin with the antiparallel, localized spins. The spinflip function 
(8) does not at all vanish as 
in the $\uparrow$-case. Nevertheless, the problem is exactly solvable resulting in a wave-vector 
independent self-energy:
\begin{equation}
\Sigma_{\downarrow}^{(T=0,n=0)}(E)
=\frac{1}{2}JS\left(1+\frac{JG_{0}(E+\frac{1}{2}JS)}{1-\frac{1}{2}JG_{0}(E+\frac{1}{2}JS)}\right).
\end{equation}
$G_{o}(E)$ is the "free" propagator:
\begin{equation}
G_{0}(E)=\frac{1}{N}\sum_{\mathbf{k}}G_{\mathbf{k}}^{(0)}(E)=\frac{1}{N}\sum_{\mathbf{k}}\frac{1}
{E-\epsilon(\mathbf{k})}.
\end{equation}

The reason for the wave-vector independence of the self-energy can be traced back [10] to the 
lack of a direct (Heisenberg) exchange term in the model-Hamiltonian (1). Therefore 
$\Sigma_{\downarrow}^{(T=0,n=0)}(E)$  does not
 contain magnon energies $\hbar\omega(\mathbf{q})$ which come into play when the excited 
 $\downarrow$-electron flips its spin by  magnon emission. Neglecting the exchange between the 
 local-moment spins $\mathbf{S}_{i}$ may be considered as the $\hbar\omega(\mathbf{q})\equiv 0$ case. As a 
 consequence, the electronic self-energy becomes wave-vector independent. There does not arise any 
 problem in calculating the limit $(n=0,\ T=0)$ with the inclusion of a Heisenberg exchange 
$(\sim J_{ij}\mathbf{S}_{i}\cdot \mathbf{S}_{j})$. Then the wave-vector dependence of the selfenergy reappears [10].

\subsection{Second-order perturbation theory}

Conventional diagrammatic perturbation theory for the Kondo-lattice model does not work 
because of the lack of Wick's theorem. A fertile alternative is the Mori-formalism [36, 37], 
which allows for a systematic expansions of the electronic selfenergy of the KLM with respect 
to the powers of $J$. That has successfully been done previously for the weakly coupled Hubbard 
model by the use of the modified perturbation theory of [38, 39]. In the case of the KLM, the first 
order term is just the mean-field result $\frac{1}{2}Jm_{\sigma}$, while in the second order, one 
finds (Eq. (3.12) in ref. 39):
$$\Sigma_{\mathbf{k}\sigma}^{(2)}(E)\stackrel{n\rightarrow 0}{\longrightarrow}
\frac{J^{2}}{4N^{2}}\sum_{\mathbf{q}}\left\{\left<S_{-\mathbf{q}}^{-\sigma}S_{\mathbf{q}}^{\sigma}
\right>^{(1)}G_{\mathbf{k+q}}^{(0)}\left(E-\frac{1}{2}Jm_{\sigma}\right)\right .$$
\begin{equation}
+\left .\left<\delta S_{-\mathbf
{q}}^{z}\cdot\delta S_{\mathbf{q}}^{z}
\right>^{(1)}G_{\mathbf{k+q}}^{(0)}\left(E+\frac{1}{2}Jm_{\sigma}\right)\right\}.
\end{equation}
$<\cdots>^{(1)}$ means mean-field averaging, while the q-dependent spin operator is defined as
 usual,
\begin{equation}
S_{\mathbf{q}}^{\alpha}=\sum_{i}S_{i}^{\alpha}e^{-i\mathbf{q}\cdot\mathbf{R}_{i}}\ (\alpha=+,-,z)
\end{equation}
 $\delta S_{\mathbf{q}}^{z}$ is a short-hand notation:
\begin{equation}
\delta S_{\mathbf{q}}^{z}=S_{\mathbf{q}}^{z}-\left<S_{\mathbf{q}}^{z}\right>^{(1)}.
\end{equation}
In the following we are interested in the local self-energy
$\Sigma_{\sigma}(E)=\frac{1}{N}\sum_{\mathbf{k}}\Sigma_{\mathbf{k}\sigma}(E)$ only, for which we 
find with (30) up to order $J^{2}$ in the limit $n\rightarrow 0$ to be :
\begin{equation}
\begin{split}
\Sigma_{\sigma}(E)=-\frac{1}{2}Jm_{\sigma}+\frac{1}{4}J^{2}\{S(S+1)\hspace{3cm}\\
-m_{\sigma}(m_{\sigma}+1)\}G_{0}(E)+O\left(J^{3}\right).
\end{split}
\end{equation}

\subsection{High-energy expansions}

For controlling unavoidable approximations, the spectral moments $M_{\mathbf{k}\sigma}^{(n)}$ of 
the spectral density $S_{\mathbf{k}\sigma}(E)$
\begin{equation}
S_{\mathbf{k}\sigma}(E)=-\frac{1}{\pi}Im G_{\mathbf{k}\sigma}(E)
\end{equation}
are of great importance:
\begin{equation}
M_{\mathbf{k}\sigma}^{(n)}=\frac{1}{\hbar}\int_{-\infty}^{+\infty}dE\cdot E^{n}\cdot
 S_{\mathbf{k}\sigma}(E).
 \end{equation}
 
In principle, they can be calculated rigorously via the equivalent expression
\begin{equation}
M_{\mathbf{k}\sigma}^{(n)}=
=\left<\left[\underbrace{\left[\cdots\left[c_{\mathbf{k}\sigma},H\right]
_{-},\cdots,H\right]_{-}}_{n-fold},c_{\mathbf{k}\sigma}^{\dag}\right]_{+}\right>.
\end{equation}

There is a close connection between the spectral moments and the high-energy behaviour of 
the Green function:
\begin{equation}
\begin{split}
G_{\mathbf{k}\sigma}(E)=\int_{-\infty}^{+\infty}dE'\frac{S_{\mathbf{k}\sigma}(E')}{E-E'}
\hspace{2.5cm}\\
=\frac{1}{E}\sum_{n=0}^{\infty}\int_{-\infty}^{+\infty}dE'\left(\frac{E'}{E}\right)^{n}S_
{\mathbf{k}\sigma}(E')\\
=\hbar\sum_{n=0}^{\infty}\frac{M_{\mathbf{k}\sigma}^{(n)}}{E^{n+1}}\hspace{3cm}.
\end{split}
\end{equation}
Because of the Dyson equation,
\begin{equation}
EG_{\mathbf{k}\sigma}(E)=\hbar+\left(\epsilon(\mathbf{k})+\Sigma_{\mathbf{k}\sigma}(E)\right)
G_{\mathbf{k}\sigma}(E)
\end{equation}
an analoguous expansion holds for the selfenergy:
\begin{equation}
\Sigma_{\mathbf{k}\sigma}(E)=\sum_{m=0}^{\infty}\frac{C_{\mathbf{k}\sigma}^{(m)}}{E^{m}}.
\end{equation}
The coefficients $C_{\mathbf{k}\sigma}^{(m)}$ turn out to be simple functions of the moments 
up to order $m + 1$:
\begin{equation}
C_{\mathbf{k}\sigma}^{(0)}=M_{\mathbf{k}\sigma}^{(1)}-\epsilon(\mathbf{k})
\end{equation}
\begin{equation}
C_{\mathbf{k}\sigma}^{(1)}=M_{\mathbf{k}\sigma}^{(2)}-\left(M_{\mathbf{k}\sigma}^{(1)}\right)^{2}
\end{equation}
\begin{equation}
C_{\mathbf{k}\sigma}^{(2)}=M_{\mathbf{k}\sigma}^{(3)}-2M_{\mathbf{k}\sigma}^{(2)}M_{\mathbf{k}
\sigma}^{(1)}+\left(M_{\mathbf{k}\sigma}^{(1)}\right)^{3}.
\end{equation}
Using the definition (36), the moments of the KLM can be explicitly calculated by the use of the 
model-Hamiltonian (1). After tedious but straightforward manipulations, one finds in the low-
density limit $(n\rightarrow 0)$, for the first four moments:
\begin{equation}
M_{\mathbf{k}\sigma}^{(0)}=1
\end{equation}
\begin{equation}
M_{\mathbf{k}\sigma}^{(1)}=\epsilon(\mathbf{k})-\frac{1}{2}Jm_{\sigma}
\end{equation}
\begin{equation}
M_{\mathbf{k}\sigma}^{(2)}=\epsilon^{2}(\mathbf{k})-Jm_{\sigma}\epsilon(\mathbf{k})+\frac{1}{4}J^{2}
(S(S+1)-m_{\sigma})
\end{equation}
\begin{equation}
\begin{split} 
M_{\mathbf{k}\sigma}^{(3)}=\epsilon^{3}(\mathbf{k})-\frac{3}{2}Jm_{\sigma}\epsilon^{2}(\mathbf{k})
+\frac{1}{4}J^{2}\{A_{\sigma}(\mathbf{k})+\\
+B(\mathbf{k})+2\epsilon(\mathbf{k})(S(S+1)-m_{\sigma})\}+\\
+\frac{1}{8}J^{3}\{S(S+1)(1-m_{\sigma})-m_{\sigma}\}.
\end{split}
\end{equation}
$A_{\sigma}(\mathbf{k})$ and $B(\mathbf{k})$ are related to spin-correlation functions:
\begin{equation}
A_{\sigma}(\mathbf{k})=\frac{1}{N}\sum_{i,j}e^{i\mathbf{k}\cdot(\mathbf{R}_{i}-\mathbf{R}_{j})}T_{ij}\left<
S_{i}^{-\sigma}\cdot S_{j}^{\sigma}\right>
\end{equation}
\begin{equation}
B(\mathbf{k})=\frac{1}{N}\sum_{i,j}e^{i\mathbf{k}\cdot(\mathbf{R}_{i}-\mathbf{R}_{j})}T_{ij}\left<
S_{i}^{z}\cdot S_{j}^{z}\right>.
\end{equation}
Inserting these expressions into eqs. (40-42) we get for the first three selfenergy 
coefficients:
\begin{equation}
C_{\mathbf{k}\sigma}^{(0)}=-\frac{1}{2}Jm_{\sigma}\hspace{4cm}
\end{equation}
\begin{equation}
C_{\mathbf{k}\sigma}^{(1)}=-\frac{1}{4}J^{2}(S(S+1)-m_{\sigma}(m_{\sigma}+1))
\end{equation}
\begin{equation}
\begin{split}
C_{\mathbf{k}\sigma}^{(2)}=-\frac{1}{4}J^{2}(A_{\sigma}(\mathbf{k})+B(\mathbf{k})-\epsilon(\mathbf{k})
m_{\sigma}^{2})+\hspace{0.5cm}\\
+\frac{1}{8}J^{3}(1+m_{\sigma})(S(S+1)-m_{\sigma}(m_{\sigma}+1)).
\end{split}
\end{equation}
They determine the high-energy behaviour of the selfenergy (39).

\subsection{Interpolation formula}

We want to construct an approximate expression for the electronic selfenergy of the low-
density KLM, which fulfills the zero-bandwidth limit (26) for all temperatures T and arbitrary 
coupling strengths $J$, as well as the exact $T=0$-result (27) and (28) for arbitrary bandwidths 
and couplings. Furthermore, it should reproduce the correct high-energy (strong coupling) 
behaviour (39) and in addition also the weak-coupling result (33). Guided by the non-trivial $(n=0, 
T=0)$-result (28), we start with the following ansatz for the local selfenergy:

\begin{equation}
\Sigma_{\sigma}(E)=-\frac{1}{2}Jm_{\sigma}+\frac{1}{4}J^{2}\frac{a_{\sigma}G_{0}(E-\frac{1}{2}
Jm_{\sigma})}{1-b_{\sigma}G_{0}(E-\frac{1}{2}Jm_{\sigma})}.
\end{equation}

$a_{\sigma}$ and $b_{\sigma}$ are at first unknown parameters. It is easy to recognize that this
 ansatz reproduces the exact limit (27) and (28) of ferromagnetic saturation, if $T=0,n=0$,
 \begin{equation}
 a_{\sigma}=(1-z_{\sigma})S\hspace{0.2cm};\hspace{0.2cm}b_{\downarrow}=\frac{1}{2}J\hspace{0.2cm}
 ;\hspace{0.2cm}b_{\uparrow}\  arbitrary
 \end{equation} 
and the zero-bandwidth limit (26), if
$$\epsilon(\mathbf{k})\rightarrow T_{0}\hspace{0.2cm}\forall\mathbf{k}$$
\begin{equation}
a_{\sigma}=S(S+1)-m_{\sigma}(m_{\sigma}+1)
\end{equation}
$$b_{\sigma}=b_{-\sigma}=\frac{1}{2}J$$
We note that (54) agrees with (53) for $\mbox{T=0}$!

By (52), we concentrate ourselves from the very beginning on the local part of the selfenergy. 
As already stated above, the wave-vector dependence of the selfenergy is mainly due to 
magnon energies $\hbar\omega(\mathbf{q})$ appearing at finite temperature in magnon emission and 
absorption processes by the band electron. However, the neglect of a direct Heisenberg exchange 
between the localized spins in the KLM can be interpreted as the
$\hbar\omega(\mathbf{q})\rightarrow 0$ limit.

We fix the parameters $a_{\sigma}$ and $b_{\sigma}$ in the ansatz (52), by equating the high-energy
 expansion (39). For this purpose, we first develop (52) in terms of powers of the inverse energy. 
 That requires the respective high-energy expression of the mean-field propagator
 $G_{0}(E-\frac{1}{2}Jm_{\sigma})$, which is exactly known:
\begin{equation}
G_{0}\left(E-\frac{1}{2}Jm_{\sigma}\right)=\sum_{n=0}^{\infty}\frac{\hat{M}_{\sigma}^{(n)}}
{E^{n+1}}
\end{equation}
\begin{equation}
\hat{M}_{\sigma}^{(n)}=\frac{1}{N}\sum_{\mathbf{k}}\left(\epsilon(\mathbf{k})+\frac{1}{2}Jm_
{\sigma}\right)^{n}.
\end{equation}
From (52) it then follows:

$$\Sigma_{\sigma}(E)
=-\frac{Jm_{\sigma}}{2}+\frac{J^{2}a_{\sigma}}{4}\sum_{m=0}^{\infty}
\frac{\hat{M}_{\sigma}^{(m)}}{E^{m+1}}
 \sum_{p=0}^{\infty}\left[b_{\sigma}\sum_{n=0}^{\infty}
\frac{\hat{M}_{\sigma}^{(n)}}{E^{n+1}}\right]^{p}$$
$$=-\frac{1}{2}Jm_{\sigma}+\frac{1}{E}\left\{\frac{1}{4}J^{2}\hat{M}_{\sigma}^{(0)}a_{\sigma}
\right\}+\frac{1}{E^{2}}\left\{\frac{1}{4}J^{2}a_{\sigma}\right .$$
\begin{equation}
\times \left . \left(\hat{M}_{\sigma}^{(1)}
 +b_{\sigma}\left(\hat{M}_{\sigma}^{(0)}\right)^{2}\right)\right\}+O\left(1/E^{3}\right).
\end{equation}
The so-derived local selfenergy coefficients,
\begin{equation}
C_{\sigma}^{(m)}=\frac{1}{N}\sum_{\mathbf{k}}C_{\mathbf{k}\sigma}^{(m)}
\end{equation}
\begin{equation}
C_{\sigma}^{(0)}=-\frac{1}{2}Jm_{\sigma}
\end{equation}
\begin{equation}
C_{\sigma}^{(1)}=\frac{1}{4}J^{2}a_{\sigma}
\end{equation}
\begin{equation}
C_{\sigma}^{(2)}=\frac{1}{4}J^{2}a_{\sigma}\left(T_{0}+\frac{1}{2}Jm_{\sigma}+b_{\sigma}\right)
\end{equation}
can be compared to the exact expressions following from (49)-(51):
\begin{equation}
C_{\sigma}^{(0)}=-\frac{1}{2}Jm_{\sigma}
\end{equation}
\begin{equation}
C_{\sigma}^{(1)}=\frac{1}{4}J^{2}(S(S+1)-m_{\sigma}(m_{\sigma}+1))
\end{equation}
\begin{equation}
\begin{split}
C_{\sigma}^{(2)}=\frac{1}{4}J^{2}\left(T_{0}+\frac{1}{2}J(1+m_{\sigma})\right)\left(S(S+1)\right .\\
\left .-m_{\sigma}(m_{\sigma}+1)\right).
\end{split}
\end{equation}
 $C_{\sigma}^{(0)}$ is identically fulfilled. Agreement for the two other coefficients is achieved 
 by setting
\begin{equation}
a_{\sigma}=S(S+1)-m_{\sigma}(m_{\sigma}+1)
\end{equation}
\begin{equation}
b_{\sigma}=\frac{1}{2}J=b_{-\sigma}.
\end{equation}
These are the same expressions as found in (54) for the special zero-bandwidth limit.
 
Inserting (65) and (66) into (52) yields a self-energy result, which is exact for
$T=0\ (m_{\sigma}=z_{\sigma}S)$  but arbitrary bandwidths $W$ and exchange couplings $J$. It 
fulfills the zero-bandwidth limit for all couplings $J$ and all temperatures $T$. It obeys the 
high-energy behaviour which is important for the strong-coupling regime. Furthermore, the 
comparison with (33) shows that the approach fits second-order perturbation theory, thus being 
reliable in the weak-coupling regime, too. We believe that (52) together with (65) and (66) 
represents a trustworthy approach to the low-density self-energy of the Kondo-lattice model. 
In the next Section we present a numerical evaluation.

\begin{figure}[htb]
\begin{center}
\hfill\includegraphics[width=0.4\textwidth]{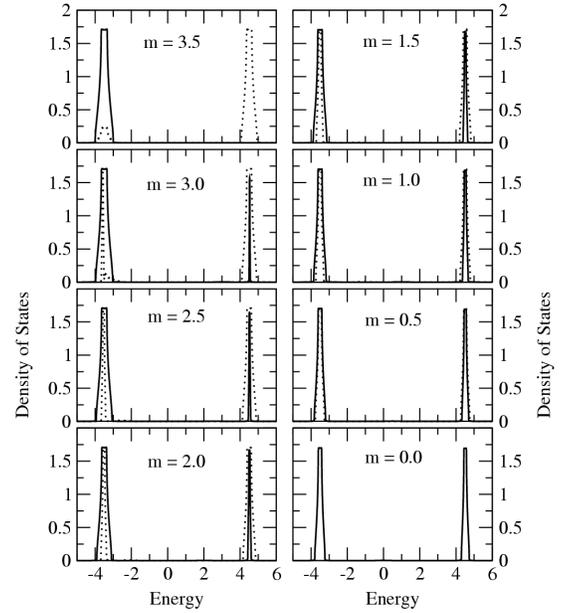}\hfill
  \end{center}    
    \caption{Quasi-particle density of states as a function of energy for various 
      values of magnetization. Full line for spin up and dotted line for spin
      down. $J = 2$, $S= 7/2$ and $W = 1$.}    
\end{figure}

\section{Results}

We have evaluated our theory for an sc lattice using the respective Bloch-density of states (B-DOS)
 in tight-binding approximation [40]. The center of gravity $T_{0}$ of the Bloch-band is 
chosen as energy zero. Fig.1 shows the temperature-dependent quasiparticle density of states 
(Q-DOS) $\rho_{\sigma}(E)$ for a strongly coupled system ($J=2.0\ eV,\ S=7/2,\ W=1\ eV$).
 The electronic spectrum gets its temperature-dependence exclusively through the local-moment 
magnetization $m=|m_{\sigma}|=\left|\left<S^{z}\right>\right|$, which must be considered as 
an external parameter. $m =3.5$ means $T=0K$ (ferromagnetic saturation), while $m=0$ belongs to 
$T=T_{c}$. The Q-DOS for each spin direction consists of two subbands separated by an energy of 
the order $\frac{1}{2}J(2S+1)$. They originate from the two atomic levels $E_{1\sigma}$ and
$E_{2\sigma}$ in the zero-bandwidth limit (16).

\begin{figure}[htb]
\begin{center}
\hfill\includegraphics[width=0.4\textwidth]{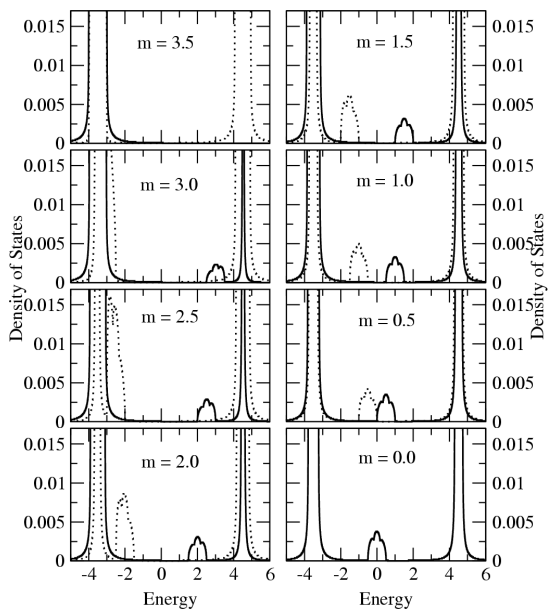}\hfill
  \end{center}    
    \caption{Same as in figure 1 but with enlarged vertical scale}    
\end{figure}

A special case is the ferromagnetic saturation, for which the $\uparrow$-spectrum consists only of
 the undeformed low-energy band ($\rho_{\uparrow}(E)=\rho_{0}(E+\frac{1}{2}JS)$). The 
 $\uparrow-$electron has no chance to exchange its spin with the perfectly 
 aligned local-spin system. The spinflip terms in the exchange interaction (4) therefore do not 
 work, only the Ising-like part (first term in (4)) takes care for a rigid shift of the excitation
  spectrum. The $\downarrow-$spectrum is more complicated because a $\downarrow$-electron can, even
   at $T=0K$, exchange its spin with the ferromagnetically saturated spin system. One possibility 
is to emit a magnon, therewith reversing its own spin and becoming a $\uparrow-$electron. Such a 
spinflip-excitation is of course possible only if there are $\uparrow-$states within reach on 
which the original $\downarrow-$electron can land after the spinflip. That is the reason why the 
low-energy $\downarrow-$subband occupies the same energy region as the $\uparrow-$band.
\begin{figure}[htb]
\begin{center}
\hfill\includegraphics[width=0.4\textwidth]{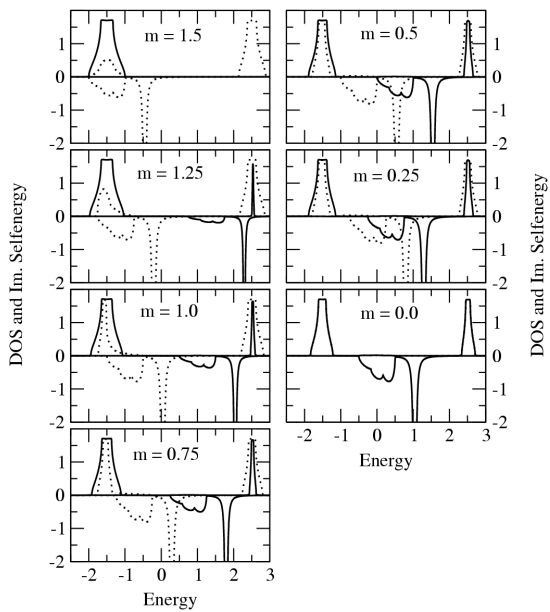}\hfill
  \end{center}    
    \caption{Quasi-particle density of states(in the positive half of the frame)
    and imaginary part of the selfenergy(in the negative half of the frame) as a
    function of energy for various values of magnetization. Full line for spin
    up and dotted line for spin down. $J = 2$, $S= 3/2$ and $W = 1$.}

\end{figure}

The $\downarrow-$electron has another possibility to exchange its spin with the ferromagnetically 
saturated moment system by a repeated magnon emission and reabsorption. In a certain sense the 
electron propagates through the lattice dressed by a virtual cloud of magnons. For the parameters 
chosen in Fig.1, this gives rise even to the formation of a stable quasiparticle, which we call 
"magnetic polaron" [10, 35]. The polaron states form, at $T=0K$, the upper $\downarrow-$ 
quasiparticle subband. It goes without saying that polaron formation is impossible for the
$\uparrow-$electron in the saturated moment system. Therefore no upper quasiparticle subband 
appears in the $\uparrow-$spectrum. This changes for finite temperatures.

\begin{figure}[htb]
\begin{center}
\hfill\includegraphics[width=0.4\textwidth]{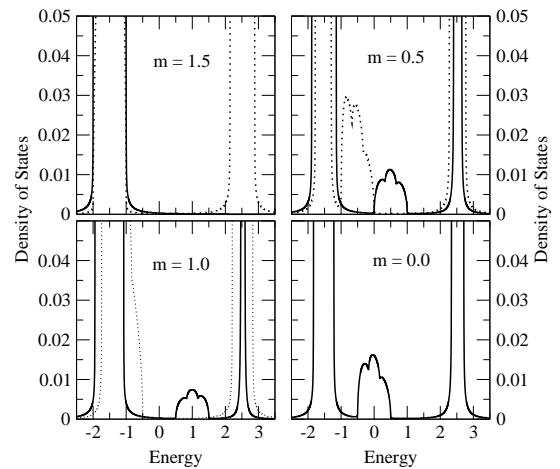}\hfill
  \end{center}    
    \caption{Quasi-particle density of states as a function of energy for various 
      values of magnetization. Full line for spin up and dotted line for spin
      down. $J = 2$, $S= 3/2$ and $W = 1$.}    
\end{figure}

For $T > 0\ (m < 3.5)$ the $\uparrow-$spectrum too, becomes more structured because the localized 
spin system is no longer perfectly aligned. The system now contains magnons which can be 
absorbed by the $\uparrow-$electron. Even polaron formation becomes possible. The spectral weight 
of the upper $\uparrow-$ quasiparticle subband raises with increasing temperature, i.e., increasing
 magnon density. Fig.1 illustrates that the temperature-dependence of the Q-DOS mainly affects the
 spectral weights of the subbands and not so much their positions. This is a typical feature of 
the strong coupling regime $JS>>W$. In such a situation, the band electron mobility is rather 
poor, it stays for a relatively long time at the same lattice site. The actual quantization axis 
is then the localized spin $(S=7/2)$, to which the electron can orient its spin parallel 
("spin up" in the local frame) or antiparallel ("spin down" in the local frame). The excitation 
energy for a parallel alignment roughly amounts to $-\frac{1}{2}JS$, and for an antiferromagnetic 
alignment, to $+\frac{1}{2}J(S+1)$. The lower quasiparticle subband consists of states 
  belonging to the situation where the band electron appears in the local frame as "spin up" 
  electron. This may happen directly or after emitting/absorbing a magnon. In the upper subband 
  the electron has entered the local frame as "spin down" electron. This is impossible for a
   $\uparrow-$electron at $T=0K$, when all localized spins are parallelly aligned $(m=S)$. While the 
 excitation energies are almost temperature-independent, the probability for the electron to be 
in the local frame as a "spin up" or as a "spin down" particle strongly depends on temperature.
 That manifests itself in the spectral weight of the respective quasiparticle subband, which 
 therefore is temperature- and spin-dependent.There remains a small probability that the band 
 electron is not trapped by the localized spin, but rather propagates with high mobility through 
 the spin lattice. In such a case the effective quantization axis is no longer the local spin but rather the direction of
 the global magnetization $\left<S^{z}\right>$. Fig.2 shows the Q-DOS for the same parameters as 
 in Fig. 1 but on a finer scale. One recognizes two tiny satellites which emerge from the two main
  peaks with increasing temperature (decreasing magnetization $m$). The $\downarrow$-satellite has
 a lower energy than the $\uparrow-$satellite. This can be understood as follows: The original
 $\downarrow-$electron, for entering the low energy part of the spectrum, will predominantly do this
 by emitting a magnon thereby reversing its own spin. In case of being not trapped by a local spin
, it then moves as a $\uparrow-$electron through the spin lattice. On the other hand, an original 
$\uparrow-$electron has to absorb a magnon in order to enter the high-energy part of the spectrum 
and propagating then as $\downarrow-$electron. With decreasing magnetization the two satellites 
collapse mean field-like. In the strong coupling regime $(JS >> W)$, pictured in Figs. 1 and 2, 
the satellites have only very small spectral weights, nevertheless representing interesting 
physics.

\begin{figure}[htb]
\begin{center}
\hfill\includegraphics[width=0.4\textwidth]{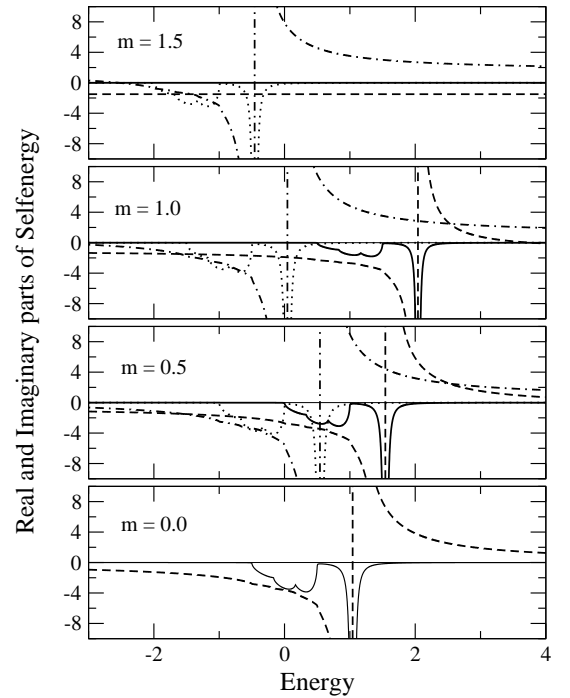}\hfill
  \end{center}    
    \caption{Real and imaginary parts of the selfenergy as a function of energy
    for different values of magnetization. For imaginary part, full line for spin up
    and dotted line for spin down. For real part, dashed line for spin up
    and dash-dotted line for spin down. Imaginary part of the selfenergy is
    multiplied by a factor of 5 for better clarity. $J = 2$, $S= 3/2$ and $W = 1$.}    
\end{figure}

The parameters used in Figs. 3, 4 and 5 $(J=2\ eV, W=1\ eV, S=3/2)$ should be typical for the 
manganites. It is sometimes claimed [21,30] that because of the strong coupling $JS$, the 
itinerant electron ($e_{g}$) spin is oriented at $T=0K$ in any case parallel to the localized 
($t_{2g}$) spin. According to the exact $m=S=3/2-$part of Fig.3, this can strictly be ruled out for the 
 FKLM. In the mentioned papers the assumption of full polarization is an artefact due to the 
restriction to classical spins ($S\rightarrow\infty$)). The temperature-dependence of the Q-DOS is
 of course very similar to the $S=7/2$ case in Fig. 1. Even the satellites, which describe the 
 "free" electron propagation after emitting/absorbing a magnon, do appear (Fig. 4). However, 
 because of the smaller distance between the two main peaks ($\approx \frac{1}{2}J (2S+1)$) the 
 mean-field shift of the satellites is not so clearly visible as for the higher spin in Fig. 1.

\begin{figure}[htb]
\begin{center}
\hfill\includegraphics[width=0.4\textwidth]{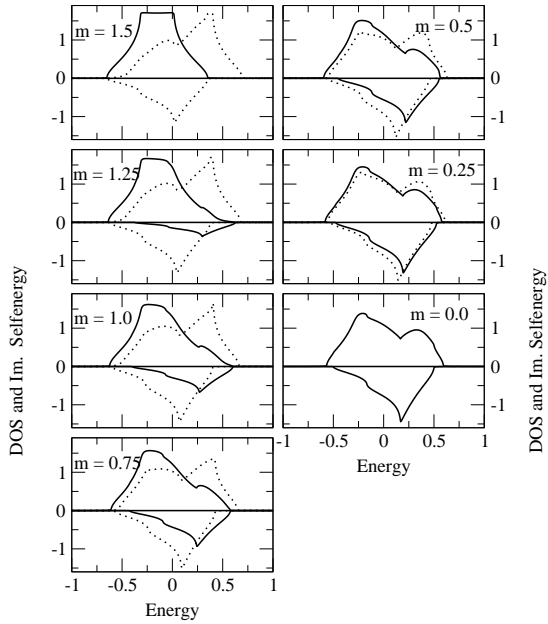}\hfill
  \end{center}    
    \caption{Quasi-particle density of states(in the positive half of the frame)
    and imaginary part of the selfenergy(in the negative half of the frame)
    as a function of energy for different values of magnetization. Full line for
    spin up and dotted line for spin down.  Imaginary selfenergy is multiplied
    by a factor of 5 for better clarity.  $J = 0.2$, $S= 3/2$ and $W = 1$.}    
\end{figure}

The imaginary part of the self-energy is directly related to quasiparticle damping and lifetime, 
respectively. Fig. 3 demonstrates that the polaron states (upper part of the spectrum) represent 
quasiparticles with almost infinite lifetimes since $Im\Sigma_{\sigma}(E)$ is zero in this region
. For $T=0K$, this is an exact result. In ferromagnetic saturation the whole $\uparrow-$spectrum
 consists of stable states. It turns out that, in the here discussed strong coupling regime, even
for finite temperatures, only the states of the mean-field satellites are getting finite lifetimes.
 The sharp peak of $Im\Sigma_{\sigma}(E)$ falls always into the bandgap, which is provoked by a 
 divergence of the real part of the selfenergy (Fig. 5). It has therefore no direct influence on 
 the lifetime of quasiparticles.
 
Up to now we have only discussed the FKLM in the strong coupling regime. As demonstrated 
in Sect. II.D, our interpolating approach is correct in the weak-coupling region, too. Fig. 6 
shows, as an example, the Q-DOS for $J=0.2eV, W=1eV$ and $S=3/2$. The tendency to the 
two-subband structure can be recognized for weak couplings, too. The physical interpretation 
of the responsible elementary processes is the same as in the above-discussed strong coupling 
case. For $T=0$ all $\uparrow-$states represent stable quasiparticles, the respective imaginary 
part of the selfenergy vanishes. With increasing demagnetization of the local moment system, 
$Im\Sigma_{\uparrow}(E)$ becomes finite indicating finite lifetimes of $\uparrow-$quasiparticles 
due to magnon absorption, which is impossible at $T=0$ because of ferromagnetic saturation. Magnon
 emission by $\downarrow-$electrons, however, is always possible. It should be pointed out that 
 the upper part of the $\downarrow-$spectrum obviously consists, at low temperature, of stable 
 polaron states.

\begin{figure}[htb]
\begin{center}
\hfill\includegraphics[width=0.4\textwidth]{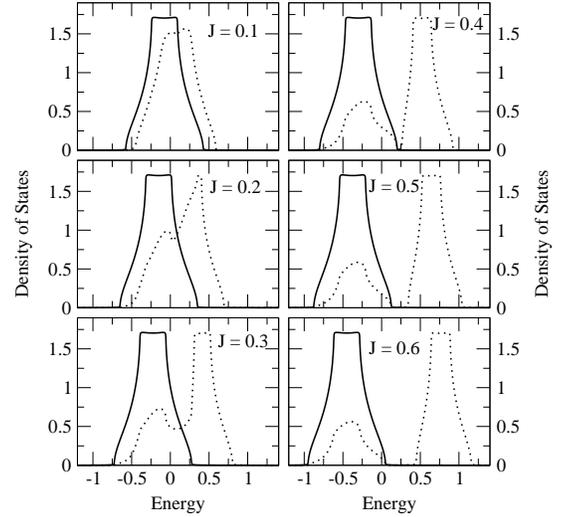}\hfill
  \end{center}    
    \caption{Quasi-particle density of states as a function of energy for
    different values of coupling constant $J$. Full line for spin up and dotted line for
    spin down. $m = 1.5$ (ferromagnetic saturation), $S= 3/2$ and $W = 1$.}    
\end{figure}

It is surprising that already very small couplings are sufficient to create a pseudo gap in the 
quasiparticle spectrum. According to Fig. 7, which shows the exact $T=0-\rho_{\sigma}(E)$ for 
various exchange couplings J, the gap opens already for $J=0.4eV$. Our results for the weakly 
coupled FKLM are very similar to those presented in ref. 10.

\section{Conclusions}

We have presented an approach to the ferromagnetic Kondo-lattice model in the low-density 
limit ($n\rightarrow 0$). The theory uses an interpolation formula for the electronic selfenergy 
which fulfills a maximum number of limiting cases. It reproduces the non-trivial rigorous special 
case of a single electron in an otherwise empty conduction band at $T=0$ (ferromagnetically 
saturated semiconductor), and that for arbitrary bandwidths and coupling constants. It is exact 
in the zero-bandwidth limit for all temperatures and all exchange couplings. It obeys the high-
energy expansion of the selfenergy, guaranteeing therewith the right strong-coupling 
behaviour, as well as perturbation theory of second order ($\propto J^{2}$) for the weak-coupling 
side. All exact criteria available for the ferromagnetic Kondo-lattice model, known to us, are 
correctly reproduced by the present low-density approach.

Strong correlation effects due to interband exchange appear in the quasiparticle density of states.
 Already a rather weak coupling $J/W$ provokes a distinct temperature-dependence in the 
electronic structure, mainly due to spin  exchange processes between the localized magnetic 
moments and itinerant band electrons. Magnon  emission/absorption processes compete with polaron-like 
quasiparticle formation. These facts demonstrate that the assumption of classical spins
($S\rightarrow\infty$), very often used for the simplified  treatment of the model [30], 
suppresses just the essentials of Kondo-lattice model. The necessary extension of the presented 
theory has to include finite band occupations, which certainly requires additional approximations.
 The here developed $n\rightarrow 0$-approach can then serve as a weighty criterion for the 
 correctness of the approach.

\section{Acknowledgements}

This work has been prepared as an India-Germany Partnership Project sponsored by the 
Volkswagen Foundation.

\end{document}